\documentclass[prl,twocolumn,nofootinbib]{revtex4-1} %{revtex4}
\usepackage{appendix}
\usepackage{anysize}
\usepackage{graphics}
\usepackage{color}
\usepackage{amssymb}
\usepackage{graphicx}
\usepackage{epstopdf}
\usepackage{hyperref}
\usepackage{amsmath}
\usepackage{latexsym}
\usepackage{setspace}

\def \beq{\begin{equation}}
\def \eeq{\end{equation}}
\def \bea{\begin{align}}
\def \eea{\end{align}}

\newcommand{\avg}[1]{\langle{#1} \rangle}

\begin{document}

%\vspace{.2in}
\title{The Sunyaev-Zeldovich Signal of the maxBCG SDSS Galaxy Clusters in WMAP}

%\vspace{.2in}
\author{Patrick Draper$^1$, Scott Dodelson$^{2,3,4}$, Jiangang Hao$^2$, Eduardo Rozo$^{3,4}$
}
%\vspace{.2in}

\affiliation{$^1$Enrico Fermi Institute, The
University of Chicago, Chicago, IL~~60637}
\affiliation{$^2$Center for Particle Astrophysics, Fermi National
Accelerator Laboratory, Batavia, IL~~60510}
\affiliation{$^3$Department of Astronomy \& Astrophysics, The
University of Chicago, Chicago, IL~~60637}
\affiliation{$^4$Kavli Institute for Cosmological Physics, Chicago, IL~~60637}
\date{\today}
\begin{abstract}
The Planck Collaboration measured 
the Sunyaev-Zel'dovich (SZ) decrement of optically selected clusters from the Sloan Digital Sky Survey,
finding that it falls significantly below expectations
based on existing mass calibration of the maxBCG galaxy clusters.
Resolving this tension requires either the data to go up, or the theoretical expectations to come down.  
Here, we use data from the Wilkinson Microwave Anisotropy Probe (WMAP) to perform an independent estimate of the SZ decrement
of maxBCG clusters.  The recovered signal is consistent with that obtained using Planck, though with larger error bars due
to WMAP's larger beam size and smaller frequency range.
Nevertheless, this detection serves as an independent confirmation of the magnitude of the effect, and 
demonstrates that the observed discrepancy must be theoretical in origin.
\end{abstract}
\maketitle

{\noindent \it Introduction.}
Galaxy clusters are the largest gravitationally bound objects in the Universe and potentially powerful probes of dark matter and dark energy. Understanding clusters well enough to use them as probes requires a multi-wavelength approach, as observations from the radio to the gamma ray end of the spectrum reveal different features of clusters. Observations of the cosmic microwave background (CMB) can detect the Sunyaev-Zel'dovich (SZ) effect \citep{sz1972,birkinshaw1999,carlstrom2002}, wherein hot electrons distort the spectrum of passing photons. The distortion has a characteristic spectral shape and the morphology of the signal traces the integrated pressure of the cluster. 

The SZ effect was first detected with high resolution and sensitivity measurements of individual clusters, and these have become so powerful that hundreds of clusters are expected to be detected by telescopes such as the South Pole Telescope (SPT) \citep{spt} and Atacama Cosmology Telescope (ACT) \citep{act}. However, with large galaxy surveys such as the Sloan Digital Sky Survey (SDSS), which detect tens of thousands of clusters optically, even relatively low resolution CMB surveys such as WMAP \citep{wmap} can detect the SZ signal by stacking many clusters with the same optical properties. 

Recently, the \citet{:2011an} performed such a measurement utilizing the SDSS maxBCG cluster catalog \cite{koester07}.  This catalog has been used to place tight cosmological constraints on the amplitude of matter fluctuations \citep{rozoetal10a} that are consistent with those derived from X-ray selected
cluster catalogs \citep{henryetal09,vikhlininetal09b,mantzetal10a}.  Moreover, the corresponding cluster mass estimates from \citet{johnston07} --- modified as per \citet{rozoetal09} --- are consistent with the X-ray luminosity as estimated in \citep{rykoff08}, and velocity dispersion measurements as estimated by
\citep{becker07}.  Surprisingly, the \citet{:2011an} found that the SZ-decrement from these clusters fell significantly below their expectations.
If real, this discrepancy signals that either some aspect of the physics of the intra-cluster medium is not properly understood, that the selection 
of optically selected clusters is not adequately understood despite the successes it has enjoyed thus far, or that there is a hidden systematic
in the data.  We note too that this discrepancy may not the same as that discussed in other works that compare the predicted and observed
SZ signal of X-ray selected clusters (e.g. \citep{lieuetal06,bielbyshanks07,diegopartridge10,Komatsu:2010fb}, though see also 
\citep{afshordietal07,melinetal11}), as \citep{planck11_xray} explicitly checked that their SZ estimates are in fact consistent with expectations from X-ray.

Here we attempt to confirm the Planck measurement by searching for the SZ signal in WMAP caused by the SDSS maxBCG galaxy clusters.
We analyze the data using both parametric and non-parametric techniques. In both cases, we detect a signal that increases with cluster richness, with an amplitude consistent with that observed by Planck~\cite{:2011an}.  Throughout this work, we adopt a fiducial flat $\Lambda$CDM cosmology 
with $\Omega_m=0.3$ and $h=0.7$.
Cluster masses are defined within a radius $R_{500}$ such that the overdensity is 500 times that of the critical density of the universe.

{\noindent \it Data.} 
From the WMAP single-year, foreground-reduced, W-band, high-resolution ($N_{side}=1024$) temperature fluctuation maps, we take the noise-weighted mean to construct 7-year coadded maps. For each year, we subtract off an overall map mean found from the noise-weighted average of $10^6$ 
random, unique, unmasked pixels containing no MaxBCG clusters and removing outliers. The resulting map contains the temperature and noise in 12.6 million $(3.4')^2$ pixels.  

We use the WMAP maps to estimate the SZ signal of maxBCG galaxy clusters \cite{koester07}.  The catalog is constructed by
searching for red-sequence galaxies drawn from the SDSS Data Release 5, and includes
13,823 galaxy clusters over 7500 deg$^2$ in a redshift slice $z\in[0.1,0.3]$. 
Each cluster is assigned a richness $N_{200}$, the number
of galaxies within $2\sigma$ of the red-sequence and within a specified radial aperture \citep[see][for details]{koester07}.
Cluster redshifts are photometrically estimated and are accurate at the $\Delta z \lesssim 0.01$
level.  The input for our analysis is the position angle ($RA$, $DEC$), redshift ($z$), and 
richness ($N_{200}$) of each cluster. 
%If two clusters lie in the same pixel, we omit the cluster with the lower $N_{200}$.

{\noindent \it Analysis.}
\newcommand\tsz{\Delta T_{\rm SZ}}
The SZ signal at angular position $\vec\theta$ relative to a cluster center is an integral of the pressure $P$ along the line of sight:
\begin{equation}
\frac{\tsz(\vec\theta;\nu)}{T} = \frac{\sigma_T g(\nu)}{m_e} \int_{-\infty}^\infty dl P\left(\vec x[\vec\theta,d_A(z),l]\right),
\label{eq:tsz}
\end{equation}
where $g$ is the characteristic spectral shape of the SZ distortion equal to $-2$ at very low frequencies, $\sigma_T$ is the Thomson cross section, and 
--- with the cluster centered at the origin --- the 3D position $\vec x$ has components equal to $(d_A(z)\vec\theta,l)$, where $d_A(z)$ is the angular diameter distance out to the redshift of the cluster. 

%We employ two complementary techniques to extract the weak SZ signal from the WMAP data: a parametric and a non-parametric approach. The former assumes a particular form for the angular profile, leaving only the amplitude of the signal free. This is the so-called {\it matched filter} \citep{Komatsu:2010fb}, and mirrors the 
%\citet{:2011an} analysis. 
%In the non-parametric case, we explicitly compute the CMB temperature profile about galaxy clusters, and measure the corresponding
%SZ decrement.
%Either approach can be quantified by fitting for the overall amplitude of the signal, and the amplitudes cross-checked for consistency. In both analyses, we restrict the cluster set to those whose centers that lie in unmasked WMAP pixels.  If two clusters lie in the same pixel, we omit the cluster with the lower $N_{200}$. 

We employ two complementary techniques to extract the weak SZ signal from the WMAP data: a parametric and a non-parametric approach. 
In the parametric (``matched filter") approach, we take as input the data $d_i=\Delta T_i$ in pixels around each cluster, the covariance matrix $C_{ij}$ connecting the pixels (which includes both instrumental noise and correlations due to the primordial CMB), and a morphological template $t_i$ for the signal in each pixel. 
We then minimize $\chi^2$ with
\begin{equation}
\chi^2(A) = \sum_{i=1}^{N_{\rm pixels}} \left( d_i - A t_i\right) C^{-1}_{ij} \left(d_j - A t_j \right)\nonumber
\end{equation}
where $A$ is the dimensionless amplitude of the signal. The sum is over all $N_{\rm pixels}$ within an angular radius 
$\theta_{max}=5\times\theta_{500}$ of the cluster center, where $\theta_{500}=R_{500}/d_A(z)$. The estimate of $R_{500}$ requires 
a mass-richness relation, which we take to be (\cite{rozoetal09})
\begin{equation}
\avg{M_{500}|N_{200}} = e^ {B_{M|N} } \left( \frac{N_{200}}{40} \right)^{\alpha_{M|N}} \times 10^{14} M_\odot,\nonumber
\label{eq:M-N}
\end{equation}
with $B_{M|N}=0.95 \pm 0.12$ and $\alpha_{M|N}=1.06 \pm 0.11$.
Table I lists the richness bins in $N_{200}$, the assumed masses, and the angular extent of the clusters at $z=0.2$.

%\medskip
\begin{table}
%\begin{center}
\squeezetable
\begin{tabular}{ @{}| l || c | c | c | c |@{}}
\hline	
$N_{200}$ bins 	& Cluster & $M_{500}$  &	$\theta_{500}$  & $\tilde{Y}_{500}\times 10^4  $\\
($\langle N\rangle$)& Count & ($10^{13} M_\odot$) & (arcmin) & (arcmin$^2$) \\
\hline
\hline
  8-30 (14) & 12408 & 8.5 & 3.2 & $0.33 \pm .23 $\\ 
  31-50 (37) & 690  & 24 & 4.5 & $2.4 \pm 1.3$ \\
  51-80 (60) & 160  & 40 & 5.3 & $4.5 \pm 2.8$ \\
81-200 (104) & 33 & 71 & 6.4 & $17 \pm 5.9$ \\
\hline  
\end{tabular}
\caption{Properties of the four bins in $N_{200}$. The number of clusters is taken after the WMAP mask is applied, and the mean number in each bin is used to calculate $M_{500}$ and $\theta_{500}$ for a redshift $z=0.2$. The SZ measurement is given in the final column.  The errors are estimated from the filter noise.} 
\label{bins}
%\end{center}
\end{table}
%\medskip

To obtain the elements of the covariance matrix, we first compute the beam-convolved\footnote{We take a Gaussian beam of variance $(0.1^{\circ})^2$ for the $W$ frequency band.} CMB correlation function $\xi(\theta)$ using the $C_l$ corresponding to WMAP7 parameters \citep{Komatsu:2010fb}. For each pair of pixels $i,j$ we determine the angular separation $\theta_{ij}\equiv [\theta_i^2+\theta_j^2-2\theta_i\theta_j\cos( \phi_i-\phi_{j}) ]^{1/2}$ between the centers of the pixels, and set the element of the covariance matrix $C_{ij}=\xi(\theta_{ij})$. The full covariance matrix is set by adding the diagonal noise contribution $N_i^2\delta_{ij}$.
For the template, we choose the pressure profile in \citet{Arnaud:2009tt}:
\begin{align}
P(r) &= 1.65\, {\rm eV}\, {\rm cm}^{-3} E^{8/3}(z) \nonumber\\
&\times \left[ \frac{ \avg{M_{500}|N_{200}} }{3\times 10^{14} M_\odot} \right]^{0.79} 
p(r/R_{500}),
\label{pressure}
\end{align}
where 
%$\avg{M_{500}|N_{200}}$ is given by equation \ref{eq:M-N}, and 
$p(x)$ is
the dimensionless profile
\begin{equation}
p(x) \equiv \frac{8.403}{(c_{500}x)^\gamma[1+(c_{500}x)^\alpha]^\delta}\nonumber
\end{equation}
with $c_{500}=1.177, \alpha=1.051,\gamma=0.3081,$ and $\delta=4.931$.
The predicted signal in a pixel $i$ is $\tsz$ as computed from 
Eq.~(\ref{eq:tsz}),
convolved with the WMAP beam $B(\vec\theta-\vec\theta_i)$:
\begin{equation}
t_i = \int d^2\theta\,B(\vec\theta-\vec\theta_i)\,\tsz(\vec\theta).\nonumber
\end{equation}

Minimization of the $\chi^2$ for a given cluster $\alpha$ leads to an estimate of the amplitude:
\begin{equation}
\hat A_\alpha=   \sum_{ij} \sigma_{A,\alpha}^2t_iC^{-1}_{ij}d_j\nonumber
\end{equation}
with ``filter noise" (the error due solely to instrument noise$+$CMB) defined as
\begin{equation}
\sigma_{A,\alpha}^{-2}=\sum_{ij} t_iC^{-1}_{ij}t_j.\nonumber
\label{noise}
\end{equation}
An amplitude estimate $\hat A_\alpha=1$ corresponds to data that is exactly consistent with the assumed template. The mean amplitude $\langle \hat A \rangle$ for a richness bin is defined as the inverse-variance weighted average of $\hat A_{\alpha}$ over all clusters in the bin. Due to intrinsic scatter in the mass-richness relation, the total error on $\hat A$ will be larger than the filter noise.
Therefore, we inverse-variance weight the amplitudes by the total error, defined as the sum in quadrature of the filter error and the intrinsic scatter.
%\begin{align}
%\Sigma_{A,\alpha}^{2}&\equiv \sigma_{A,\alpha}^2 + N\sigma_{\langle A\rangle,intrinsic}^2\;,
%\end{align}
%where $N$ is the number of clusters in the bin, and
To estimate the intrinsic scatter, we compute the scatter in a given richness bin,
\begin{equation}
%\begin{align}
%\sigma_{\langle A\rangle,intrinsic}^2&\equiv \sigma_{\langle A\rangle,sample}^2-\sigma_{\langle A\rangle,filter}^2\;,\nonumber\\
%\sigma_{\langle A\rangle,filter}^2&\equiv \left(\sum_{i=1}^{N}\sigma^{-2}_{A,i}\right)^{-1}\;,\nonumber\\
\sigma_{\langle A\rangle,total}^2 =
%&\equiv
\frac{1}{N(N-1)}\sum_{i=1}^{N}(\hat A_i-\langle \hat A\rangle)^2\; \nonumber
\label{sigmas}
\end{equation}
%\end{align}
and then subtract $\sigma_{\langle A\rangle,filter}^2$. Since the weighting depends on the total error, we iterate until the input total error is equal to the output scatter.
%Here $\sigma_{\langle A \rangle,total}$ estimates the total dispersion in the mean estimate $\langle \hat A \rangle$. Since $\Sigma_{A,\alpha}$ depends on $\langle \hat A \rangle$ through $\sigma_{\langle A\rangle,sample}$, it is necessary to begin with the approximation $\Sigma_{A,\alpha}\approx\sigma_{A,\alpha}$, and then iterate the evaluation of $\langle \hat A \rangle$ and $\Sigma_{A,\alpha}$ a few times until convergence is achieved.
%
%As given in Eq.~(\ref{sigmas}), we can estimate the contribution to the scatter in the mean from intrinsic scatter by taking the quadrature difference of the sample standard deviation over $\sqrt{N}$ and the filter noise contribution $\sigma_{\langle A \rangle, filter}$.

The template and the measured value for the amplitude $A$ can be transformed into an estimate of any related quantity. One relevant quantity
%Two relevant quantities are $Y_{500}$ and 
is $\tilde{Y}_{500}$, defined as
\begin{equation}
%\begin{align}
\tilde{Y}_{500} 
%&
\equiv \frac{\sigma_T}{m_ed_A^2}E^{-2/3}(z)\left(\frac{d_A(z)}{500\,{\rm Mpc}}\right)^2\int d^3r P(\vec r) \nonumber
%,\nonumber\\
%\tilde{Y}_{500}&\equiv Y_{500}E^{-2/3}(z)\left(\frac{d_A(z)}{500\,{\rm Mpc}}\right)^2
%\label{Ytilde}
%\end{align}
\end{equation}
where the integral is restricted to $r<R_{500}$.
Using the template in Eq.~(\ref{pressure}), we find 
 \begin{equation}
\tilde{Y}_{500}  = 8.5\times 10^{-4} A  
 \left( \frac{ \avg{M_{500}|N_{200}} }{3\times 10^{14} M_\odot}\right)^{1.79}
\,{\rm arcmin}^2.\nonumber
\label{eq:modely500}
\end{equation}

We also carry out a non-parametric fit by averaging the temperatures in pixels binned into annuli around the clusters.
As in the matched filter approach, the covariance matrix for the annuli-averaged temperatures contains contributions from instrument noise and from the primordial CMB. The first is straightforward to propagate using the combined weights from the single year WMAP maps. To compute the second, we first note that the estimated temperature in a given annulus a distance $\theta$ from a given point, taken to be the origin, is $
%\begin{equation}
T_\theta = \frac{1}{N}\sum_{i=1}^N T(\theta,\phi_i),
$%\end{equation}
 where the sum is over all $N$ pixels in the annulus, each identified by its polar position $\phi_i$. The covariance matrix between measurements in two annuli at $\theta$ and $\theta'$ is
\begin{equation}
C_{\theta\theta'} = \frac{1}{NN'} \sum_{i=1}^N\sum_{i'=1}^{N'} \langle T(\theta,\phi_i) T(\theta',\phi'_{i'}) \rangle.\nonumber
\end{equation}
Here $N'$ is generally not equal to $N$ since larger annuli have more pixels (of fixed size) in them.
The average of the product of the temperatures is $\xi(\theta_{ii'})$, the correlation of the CMB between two pixels separated by angular distance $\theta_{ii'}$.

We use 11 annuli with angular distance from the cluster equal to $(0.5,0.9,1.3,\ldots,4.5)\times \theta_{500}$.
The CMB noise is the same order of magnitude as the instrumental noise but is highly correlated 
between rings, so a simple $\chi$-by-eye of the profile is not sufficient to interpret 
the results. We ran 100 mocks to test this approach and found that the extracted amplitude $A$ is
biased high by $1.35\pm0.25$ for the $51-80$ richness bin and $1.67\pm 0.62$ for the top bin.  We correct for these biases below.%in the values we report.

\begin{figure}[t]
\begin{center}
\includegraphics[width=\columnwidth,trim=0.0in 0.3in 0.4in 0.3in,clip=true]{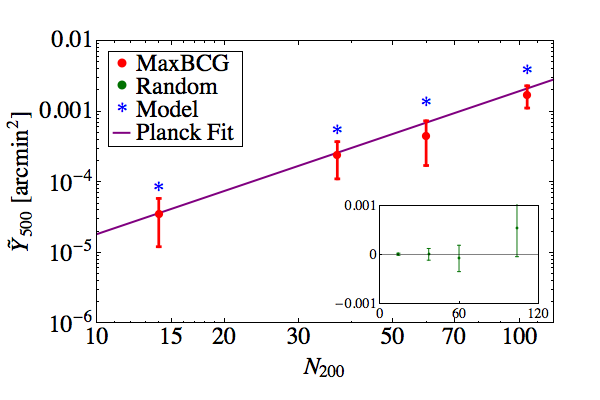}
\end{center}
\caption{W band SZ signal in 4 $N_{200}$ bins. The blue stars show the expectation for the SZ decrement of the maxBCG clusters
based on the Rozo et al. (2009) mass-richness relation, 
while the red circles show the signal measured using WMAP data.  The error bars are set by the filter noise. 
%The dashed red line gives the best power-law fit to the data.
The solid line shows the best-fit result from Planck. The inset shows that randomizing the positions of the clusters within the survey area leads to a signal consistent with zero.
}
\label{y500}
\end{figure}

{\noindent \it Results.} 
Fig.~\ref{y500} shows our estimates for $\tilde{Y}_{500}$ for clusters in the four richness bins. 
The corresponding amplitudes are $\hat A=0.37\pm0.26, 0.42\pm0.23, 0.32\pm0.20, 0.43\pm0.15$.
The error bars are $\sigma_{\langle A\rangle,filter}$ propagated from $\hat A$ to $\tilde Y_{500}$.  We also estimate the total scatter with $\sigma_{\langle A\rangle,total}$, which is sensitive to outlier clipping. In Table~\ref{err} we give values for $\sigma_{\langle A\rangle,filter}$ and $\sigma_{\langle A\rangle,total}$ without clipping, and compute the contribution from intrinsic scatter.

%Also plotted in Fig.~\ref{y500} is the result of a fit of the unbinned data to the power law $\tilde{Y}_{500}=Y_{20}(N_{200}/20)^\alpha$. We find $Y_{20}=5\pm3(\rm{stat})\times 10^{-5}$~arcmin$^2$ and $\alpha=2.2\pm0.4(\rm{stat})$, with $\chi^2/dof=0.87$. For comparison we plot the corresponding fit from the Planck Collaboration.

Our measurements are in excellent agreement with those of the \citet{:2011an}, and are clearly lower than the theoretical
predictions.  Given this independent confirmation of the tension first observed in~\cite{:2011an}, it is clear that
either the intra-cluster properties of galaxy clusters are not adequately understood, or that the optical cluster selection
suffers from a systematic that has gone undetected.  We do not currently know the solution to this problem, and indeed,
the discrepancy is likely to remain an active field of research in the immediate future.  We note that a similar discrepancy
appears to arise for the SZ signal about Luminous Red Galaxies (LRGs) \citep{handetal11}.

We perform two systematics cross-checks of our analysis pipeline: first, we randomize the location of every cluster in the catalog and repeat our 
measurement.   As expected, we find no signal.  In addition, we generate Monte Carlo realizations 
of the experiment to test whether the recovered amplitude is biased.   We randomly sample the maxBCG catalog, and assign to every cluster an SZ 
decrement (including scatter, for which we vary $\sigma_{\ln Y}$ from $0.6$ to $1.2$).  
The cluster signal of Eq.~(\ref{eq:tsz}) is convolved with the beam and added to randomly
generated (beam-convolved) CMB maps. We then fit for the amplitude of the signal using our pipeline, finding that the recovered
signal is biased low by $2\%$ ($3\%$) in the $51-80$ bin and $5\%$ ($6\%$) in the $81-200$ bin for $\sigma_{\ln Y}=0.6$ ($1.2$).

%\medskip
\begin{table}
\begin{center}
\begin{tabular}{ @{}| l || c | c | c |@{}}
\hline	
$\langle N_{200}\rangle$ & $\sigma_{\langle A\rangle,total}$  & $\sigma_{\langle A\rangle,filter}$ & $ \sigma_{\langle A\rangle,intrinsic}$\\
\hline
\hline
  14  & 0.60 & 0.23 & 0.55\\
  37 & 1.8 & 1.3 & 1.2\\
  60  & 4.3 & 2.8 & 3.2\\
 104 & 11 & 5.9 & 9.3\\
\hline  
\end{tabular}
\caption{Mean $\tilde{Y}_{500}$ errors in units of $10^{-4} $ arcmin$^2$. The filter noise includes only uncertainties from the instrument and the CMB. The quadrature difference between the total dispersion and the filter noise gives an estimate for the error in the mean due to intrinsic scatter.}
\label{err}
\end{center}
\end{table}
%\medskip

\begin{figure}[!ht]
\begin{center}
\includegraphics[width=.9\columnwidth]{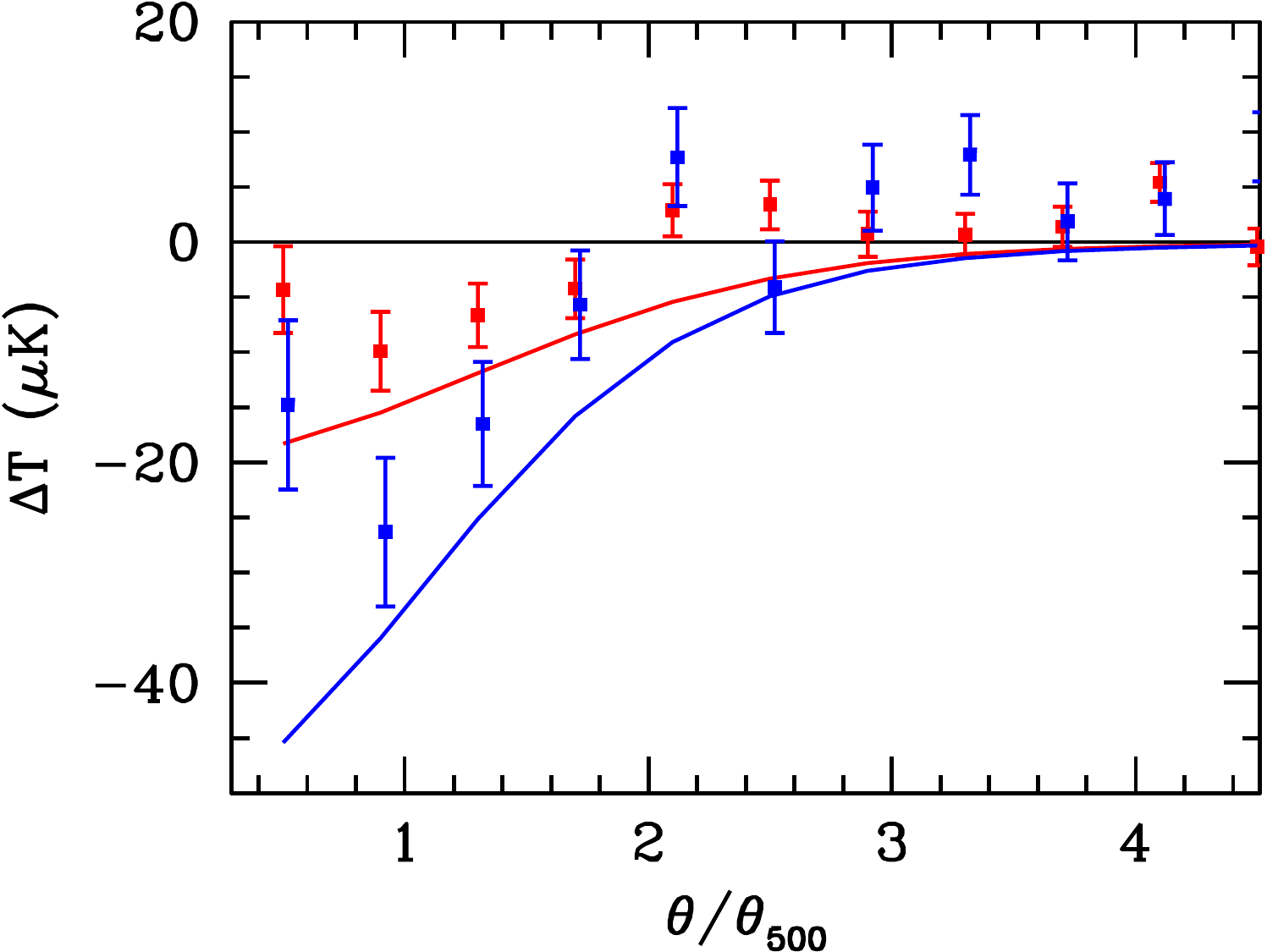}
\caption{W band SZ signal as a function of distance from the cluster center for clusters in 
the two largest richness bins listed in Table 1. Error bars include instrumental noise only. 
Solid curves are the predictions from the template assumed in Eq.~(\ref{pressure}). 
The deviation on small angular scales translates into a smaller amplitude than assumed, 
in agreement with the Planck results.}
\label{profile}
\end{center}
\end{figure}

Fig.~\ref{profile} shows the non-parametric measurement for the signal as a function of 
distance from the cluster center in the two largest richness bins. A simple $\chi^2$ 
assuming no signal reveals strong detections in both bins ($\chi^2=22.3,24.2$ for 
10 degrees of freedom each). 
The smooth curves show the predicted signal from our fiducial template
template ($A=1$). The best fit for the two bins yields 
$A=0.67\pm 0.22$ for the $N_{gals}=51-80$ bin and $A=0.67\pm0.20$ for the highest richness bin. Correcting
for the bias in these bins leads to estimates of $A=0.50$ and $0.40$ respectively, consistent within errors with 
the parametric determination. The latter is the superior 
way to determine the amplitude as it weights all pixels optimally, albeit at the cost of being sensitive
to the assumed template. 

\acknowledgements SD and JH are supported by the
US Department of Energy, including grant DE-FG02-95ER40896; and SD by National Science Foundation Grant AST-
0908072.  ER is funded by NASA through the Einstein Fellowship Program, grant PF9- 00068.

\end{document}